\documentclass[aps, twocolumn, letterpaper, showpacs, prl]{revtex4}

\usepackage{amsmath}
\usepackage{amssymb}
\usepackage{mathrsfs}
\usepackage{xspace}
\usepackage{graphicx}

\usepackage{ifpdf}
\ifpdf
\fi

\newcommand{\eg}{{e.g.,\/}\xspace}
\newcommand{\ie}{{i.e.,\/}\xspace}

\newcommand{\Eq}[1]{Eq.~(\ref{#1})}

\newcommand{\Ref}[1]{Ref.~\cite{#1}}
\newcommand{\Refs}[1]{Refs.~\cite{#1}}

\newcommand{\mc}[1]{\mathcal{#1}}

\newcommand{\pd}{\partial}

\newcommand{\msection}[1]{\textit{#1.}\ ---\ }

\begin{document}

\title{On the nature of kinetic electrostatic electron nonlinear (KEEN) waves}

\author{I.~Y. Dodin and N.~J. Fisch}
\affiliation{Princeton Plasma Physics Laboratory, Princeton University, Princeton, New Jersey 08543, USA}

\begin{abstract}
An analytical theory is proposed for the kinetic electrostatic electron nonlinear (KEEN) waves originally found in simulations by Afeyan \textit{et~al} [arXiv:1210.8105]. We suggest that KEEN waves represent saturated states of the negative mass instability (NMI) reported recently by Dodin \textit{et~al} [Phys. Rev. Lett. {\bf 110}, 215006 (2013)]. Due to the NMI, trapped electrons form macroparticles that produce field oscillations at harmonics of the bounce frequency. At large enough amplitudes, these harmonics can phase-lock to the main wave and form stable nonlinear dissipationless structures that are nonstationary but otherwise similar to Bernstein-Greene-Kruskal modes. The theory explains why the formation of KEEN modes is sensitive to the excitation scenario and yields estimates that agree with the numerical results of Afeyan \textit{et~al}. A new type of KEEN wave may be possible at even larger amplitudes of the driving field than those used in simulations so far.
\end{abstract}


% 52.35.Mw -- Nonlinear phenomena: waves, wave propagation, and other interactions
% 52.35.Sb -- Solitons; BGK modes 
% 52.35.Fp -- Electrostatic waves and oscillations (e.g., ion-acoustic waves) 

\maketitle

\bibliographystyle{brief}

\msection{Introduction} As originally shown in \Ref{ref:bernstein57}, collisionless plasmas can support stationary nonlinear waves, commonly known today as Bernstein-Greene-Kruskal (BGK) modes. Resonant particles in such modes are trapped and phase-mixed, so Landau damping is suppressed \cite{foot:itervar}. On the other hand, trapped particles are known to be responsible for a number of instabilities \cite{foot:inst}, so BGK waves are not necessarily attractor states, and, as such, are not always easily accessible \cite{ref:demeio90}. It was shown in \Refs{tex:afeyan03, ref:johnston09} that, when excited by a strong enough force, plasma oscillations can instead saturate in the form of structures that, unlike BGK modes, are nonstationary in any frame of reference and yet are undamped too. Such modes are believed to have no fluid or linear analogs and, in one-dimensional electron plasmas (to which our discussion will be limited for clarity), were termed kinetic electrostatic electron nonlinear (KEEN) waves \cite{tex:afeyan03, ref:johnston09}.

KEEN waves were numerically observed near the branch of the dispersion relation corresponding to the electron-acoustic waves (EAW), \ie at frequencies close to $\omega_{\rm EAW} \approx 1.31 kv_T$; here $k$ is the wave number, and $v_T$ is the electron thermal speed \cite{ref:holloway91, ref:montgomery01, ref:valentini06, ref:valentini12}. (Albeit strongly damped in Maxwellian plasma, and thus rarely taken into account, EAW can be nondissipative if the particle distribution is flat at velocities close to $\omega_{\rm EAW}/k$. This occurs naturally when plasma is driven externally at frequency $\omega \approx \omega_{\rm EAW}$ for a long enough time. Similar ion waves were also discussed in \Ref{ref:valentini11}.) However, KEEN modes are qualitatively different from EAW, as they contain multiple pronounced phase-locked harmonics. The advanced numerical modeling reported recently in \Refs{ref:cheng13, tex:mehrenberger13, tex:sonnendrucker12} corroborate that such a spectrum is a robust feature of KEEN waves. In particular, it was proposed in \Ref{ref:cheng13} that KEEN waves represent essentially a superposition of BGK-like structures. One may also attribute them as ``BGK waves within BGK waves'' \cite{foot:bgk}. However, the physical nature of these structures, as well as the sensitivity of KEEN waves to the excitation scenario and the driver amplitude \cite{tex:afeyan03}, are yet to be understood in detail.

The purpose of this brief note is to offer a qualitative explanation of these issues by pointing to the connection between KEEN waves and the negative mass instability (NMI) that was recently identified for BGK-like waves in \Ref{my:nmi}. In essence, the NMI causes trapped electrons to bunch into macroparticles, which then produce sideband oscillations of the wave field, shifted from the main wave by, roughly, integers of the bounce frequency. These sidebands survive in the long run only if they are phase-locked to the main wave. This requires, for parameters at which KEEN waves have been studied so far, that the bounce frequency be somewhat higher than half of $\omega_{\rm EAW}$. Below, we explain this in detail.

\msection{Action distribution} Suppose, as in \Ref{tex:afeyan03}, that electron oscillations are excited by an external driving force with some frequency $\omega$, wave number $k$, and spatially homogeneous amplitude. Assuming that the driver is turned on slowly, both trapped and passing particles conserve certain adiabatic invariants that can be expressed in terms of their actions, $J$. The action is defined as the appropriately normalized \cite{my:actii} phase space area encircled by the particle trajectory in the frame where the driver field is stationary; this frame, $\bar{\mc{K}}$, travels with respect to the laboratory frame $\mc{K}$ at the driver phase velocity, $u = \omega/k$ (assumed nonrelativistic). For a trapped particle, the invariant is $J$ itself, whereas for a passing particle the invariant is the oscillation-center canonical momentum, $P = mu + k J\,\text{sgn}\,(v-u)$, where $m$ is the electron mass, and $v$ is the electron velocity \cite{my:acti, foot:lindberg}. 

Let us assume that both $\omega$ and $k$ are constant; then conservation of $P$ implies conservation of $J$ for passing particles too \cite{foot:Jcons}. But $J$, if normalized appropriately \cite{my:actii}, is conserved also when a particle crosses the separatrix, albeit with worse-than-exponential accuracy \cite{ref:best68, ref:timofeev78, ref:cary86}. Therefore, the action distribution, $F(J)$, is conserved throughout the entire process of the wave excitation. This gives \cite{my:actii, my:bgk}
\begin{gather}\label{eq:F}
F(J) = (k/m)[f_0(u + kJ/m) + f_0(u - kJ/m)],
\end{gather}
where $f_0(v)$ is the initial velocity distribution. The separatrix action is $J = (4/\pi) m\Omega_0/k^2$ \cite{my:actii}, where $\Omega_0 = (eEk/m)^{1/2}$ is the maximum bounce frequency, and $E$ is the amplitude of the total electric field, including both the driver and the induced field. (We assume, for clarity, that $eE > 0$ and $k > 0$.) Hence, if $E$ is small enough, the trapped distribution can be approximated with the second-order Taylor expansion of \Eq{eq:F}. The terms linear in the trapped-particle action, $\pm (k/m)^2Jf'_0(u)$, mutually cancel, out so one gets
\begin{gather}
F_t(J) \approx (k/m)[2f_0(u) + (kJ/m)^2 f''_0(u)],
\end{gather}
regardless of the value of $f'_0(u)$. For $\omega \approx \omega_{\rm EAW}$ in Maxwellian plasma assumed here, one has $f''_0(u) > 0$. Thus, such a distribution is \textit{inverted}, $F_t'(J) > 0$, as is also seen directly in simulations \cite{foot:more}. 

\msection{Instability mechanism} As shown in \Ref{my:nmi}, inverted $F_t(J)$ can be unstable due to the particle bounce frequency $\Omega(J)$ being a decreasing function of~$J$. [Note that, unlike $F_t'(J)$, the slope of the ``spatially averaged velocity distribution'' is not directly linked to trapped-particle instabilities, contrary to what is often assumed in literature.] This is explained as follows. Consider a pair of electrons bouncing in the wave potential, \ie rotating in phase space around a local equilibrium. Through Coulomb repulsion (strictly speaking, via collective fields), the leading particle increases its energy; then it moves to an outer phase orbit and slows down its phase space rotation (as $\Omega' < 0$), whereas the trailing particle moves to a lower orbit and speeds up, correspondingly. This way, mutually repelling electrons can undergo phase-bunching, or condensation, as if they had negative masses. The condensation may or may not eventually saturate in the form of a stable macroparticle, but its very formation constitutes a fundamental instability in itself. By analogy with similar effects in accelerators \cite{foot:landau} and ion traps \cite{ref:strasser02, ref:pedersen01}, the term NMI was coined for this instability in \Ref{my:nmi}. 

Now consider the effect of macroparticles as production of \textit{sidebands} of the wave field. As the driver continues to feed the instability, these sidebands grow and initiate stochastization of electron orbits in the resonance region. (One can view this as an effect akin, if not identical, to quasilinear diffusion.) The stochastization tends to flatten the trapped distribution and thus eventually suppresses the NMI. Most particles then phase-mix (cf. \Ref{ref:smith78}), so a standard, albeit non-sinusoidal \cite{foot:sin}, BGK mode is formed. However, the scenario is different when the sidebands are in approximate resonance with the main wave (and, thus, with the driver too). In that case, the system is close to periodic, so one can expect formation of invariant tori in the particle phase space, even for a relatively strong driver. Then the system can sustain large stable macroparticles and the corresponding well-pronounced sidebands phase-locked to the main wave; cf. \Refs{ref:smith78, ref:tennyson94, ref:blaskiewicz04}. Once phase-locked, the wave should also be able to tolerate moderate variations of the wave amplitude from the exact resonance, as in a typical autoresonance \cite{ref:friedland09}, without abrupt modifications of the spectrum; \ie one can expect that its nonlinear features are \textit{robust}. (But, of course, large enough variations of the wave parameters destroy the resonance.)

\msection{Phase-locking conditions} The condition of phase locking is derived as follows. Suppose the system is stabilized in a state where there are $N$ identical macroparticles per island. Assuming the trapped density is small enough, this should cause oscillations of the electrostatic potential $\varphi$ in the frame $\bar{\mc{K}}$ at harmonics of the frequency $\bar{\omega} = N\Omega(J)$, where $J$ is the characteristic action corresponding to the macroparticle trajectory. Since the driver field is stationary in $\bar{\mc{K}}$ but renders the background spatially periodic, each of these harmonics is a Bloch-Floquet wave, so the total potential is representable as $\varphi = \sum_{\ell n} \varphi_{\ell n} e^{- i\ell \bar{\omega}t + i n k\bar{x}}$. Here we assumed (as is dictated also by specific boundary conditions commonly adopted for simulations \cite{my:nmi}) that the wave is spatially periodic with the same period as the driver, and the coordinate $\bar{x}$ relates to the coordinate $x$ in $\mc{K}$ as $\bar{x} = x - ut$; hence $\varphi = \sum_{\ell n} \varphi_{\ell n} e^{- i(\ell \bar{\omega} - n \omega)t + i n k x}$. On the other hand, the temporal period of a phase-locked wave must also be the same as that of the driver. Then $\bar{\omega} = \bar{N}\omega$, or $\Omega = \bar{N}\omega/N$, where $\bar{N}$ is some integer.

Since we are interested in weakly nonlinear waves, we will assume $\Omega < \omega$, or $\bar{N} < N$. Also, noticeable nonlinear structures can be expected only at resonances of not-too-high order, \ie $\bar{N} + N$ cannot be too large. This limits KEEN waves to, say, $N \lesssim 3$ and $\bar{N} = 1$. Hence the following picture is suggested for wave excitation in initially-quiescent plasma. Assuming $\Omega$ grows starting from zero, it passes several resonances of the type
\begin{gather}\label{eq:Om}
\Omega = \omega/N,
\end{gather}
with consequently decreasing $N$. As the bounce frequency is $J$-dependent, \Eq{eq:Om} can be satisfied for more than one $N$ for a given driver. On the other hand, resonances unavoidably compete when they enter the nonlinear stage. What survives is always the strongest resonance, \ie the one that has the lowest order allowed by \Eq{eq:Om}, $N_m = \omega/\Omega_0$. Using the dimensionless variables $\kappa = kv_T/\omega_p$ and $a = eE/(m\omega_p v_T)$, where $\omega_p$ is the plasma frequency, one can express $N_m$ as follows:
\begin{gather}\label{eq:N}
N_m = (u/v_T) (\kappa/a)^{1/2}.
\end{gather}

For $u/v_T = 1.31$ and $\kappa = 0.26$, which are typical for KEEN-wave simulations, \Eq{eq:N} becomes $N_m = 2(a/a_c)^{-1/2}$, where $a_c \approx 0.11$. This shows that, at $a > a_c$, phase-locking is possible into a resonance with $N = 2$, which corresponds to two macroparticles per island. In contrast, at $a < a_c$, phase-locking is possible only at $N = 3$, which corresponds to three macroparticles per island. In the latter case, the macroparticle size is much smaller, so one can expect an abrupt modification of the wave spectrum at $a \approx a_c$. This is indeed what is seen in simulations \cite{tex:afeyan03}. Moreover, the typical KEEN mode shown in Fig.~1 of \Ref{tex:afeyan03} clearly shows the presence of exactly two macroparticles in a trapping island. 

One can also anticipate a similar threshold at $a \sim 0.5$, when \Eq{eq:N} predicts $N_m = 1$. A single macroparticle can form then and bounce resonantly to the main wave. At such large amplitudes, however, the electron quiver speed becomes comparable to $v_T$, so the above estimates (which rely on the weak-interaction model and the EAW dispersion being linear) may lack quantitative accuracy. 

\msection{Conclusions} In this brief note, we propose a basic semi-quantitative theory of KEEN waves. We argue that key to the KEEN mode formation is a specific instability, the NMI \cite{my:nmi}, that produces macroparticles out of trapped electrons. These macroparticles can, under certain conditions, become phase-locked to the main wave. For parameters typical for KEEN-wave simulations reported in literature, this requires that the bounce frequency be higher than half of $\omega_{\rm EAW}$, imposing a lower limit on the driver amplitude. This picture readily explains why the formation of KEEN modes is sensitive to the excitation scenario; \eg pre-flattening of the resonant distribution would eliminate the source of the NMI, so macroparticles would not form, and the wave would remain in the linear regime. We also propose numerical estimates that agree with existing simulation results and argue that a new type of KEEN waves may be possible at even larger amplitudes of the driving field than those tried in simulations so far. Finally, due to the general nature of the mechanism considered here, similar arguments may apply to other kinetic waves too, such as the ion-bulk waves introduced in \Ref{ref:valentini11}.

The work was supported by the U.S. DOE through Contract No. DE-AC02-09CH11466, by the NNSA SSAA Program through DOE Research Grant No. DE274-FG52-08NA28553, and by the U.S. DTRA through Research Grant No. HDTRA1-11-1-0037.

%\bibliography{main,foot}

\begin{thebibliography}{10}
\bibitem{foot:bgk}
Contrary to the claims that KEEN waves cannot be explained in terms of BGK modes, this is simply the definition of stable nondissipative nonlinear structures.
\bibitem{foot:Jcons}
This is not exactly true when, as we will assume, a driver is resonant to a
  plasma eigenmode, as the latter may exhibit a time-dependent nonlinear
  frequency shift \cite{foot:itervar}. However, this does
  change the qualitative picture presented here.
\end{thebibliography}

\noindent
\textsl{{Notice:} This manuscript has been authored by Princeton University under Contract Number 
DE-AC02-09CH11466 with the U.S. Department of Energy. The publisher, by accepting the article for publication acknowledges, that the United States Government retains a non-exclusive, paid-up, irrevocable, world-wide license to publish or reproduce the published form of this manuscript, or allow others to do so, for United States Government purposes.}

\end{document}